# Thermophoretic MHD Flow and Non-linear Radiative Heat Transfer with Convective Boundary Conditions over a Non-linearly Stretching Sheet

Shalini Jain[1] and Rakesh Choudhary[2]

[1] *Manipal University Jaipur, Jaipur, Rajasthan, 303007, India*
[2] *Manipal University Jaipur, Jaipur, Rajasthan, 303007, India*

*Email: shalini.jain@jaipur.manipal.edu*

## ABSTRACT

The effects of MHD boundary layer flow of non-linear thermal radiation with convective heat transfer and non-uniform heat source/sink in presence of thermophortic velocity and chemical reaction investigated in this study. Suitable similarity transformation are used to solve the partial ordinary differential equation of considered governing flow. Runge-Kutta fourth fifth order Fehlberg method with shooting techniques are used to solved non-dimensional governing equations. The variation of different parameters such as thermophoretic parameter, chemical reaction parameter, non- uniform heat source/sink parameters are studied on velocity, temperature and concentration profiles, and are described by suitable graphs and tables. The obtained results are in very well agreement with previous results.

## 1. INTRODUCTION

The study of MHD boundary layer flow over a stretching sheet have a great significance due to its applications in engineering and industrial field. The presence of MHD in thermal management system plays a wide role in boundary layer flow. Magnetohydrodynamics stagnation flow of a micropolar fluid embedded in a porous medium in boundary layer flow studied by Nadeem et al. [1]. MHD transient flow and heat transfer of dusty fluid in channel with variable physical parameters and Navier slip boundary condition analyzed by Makinde and Chinyoka [2]. Many authors recently investigated MHD boundary layer flow in respects of different characteristics such as Hayat and Mehmood [3], Jha and Apere [4], Giressdha et al. [5,6], Jain and Choudhary [7] and Naramgri and Soluchana [8].

Many studies related to heat and mass transfer, we are neglected Dufour and Soret effects due to its smaller order of magnitude. This type of effects are arise when density difference is present in the flow system. Soret and Dufour effects are observed as the second order phenomena and useful in fields such as geosciences, petrology, hydrology etc. Hayat et al. [9] investigated Soret and Dufour effects in 3D flow with chemical reaction and convective condition in Maxwell fluid. In extend work Hayat et al. [10] studied Soret and Dufour effects in MHD boundary layer flow peristalsis of pseudoplastic nanofluid with chemical reaction. 3D boundary layer flow of a viscoelastic nanofluid with Soret and Dufour effects examined by Ramzan et al. [11]. MHD slip flow with Soret-Dufour effects in an exponentially stretching inclined sheet were investigate by Sravanthi [12]. In this study they used to solve and find out the numerical result by Homotopy analysis solution. Recently, Zaidi and Mohyud [13] discussed analysis Soret, Dufour and chemical reaction effects of wall jet flow in the presence of MHD with uniform suction/injection.

The convective heat transfer and non-linear radiation are very effective in thermal energy storage process, nuclear plants, gas turbines etc. In recent years many authors shows their interest in non-linear thermal radiation and convective boundaries conditions. A decade ago Vajravelu [14] find the viscous flow over a non-linear stretching sheet. Several authors including Pop et al. [15], Cortell [16], Aziz [17], Hayat and Qasim [18] and Afzal [19] studied non-linear stretching sheet and convective boundary conditions during their research. Makinde et al. [20] visualized a similarity solution of MHD flow of heat and mass transfer over a vertical plate with a convective surface boundary condition. They [21] extended their previous work and studied MHD mixed convection embedded in a porous medium with a convective boundary condition from a vertical plate. They have also discussed [22] boundary layer flow of a nanofluid with a convective boundary condition in a stretching sheet. Hayat et al. [23] investigated steady flow of an Eyring Powell fluid with convective boundary conditions over a moving surface. Authors such as Akbar et al. [24], Pantokratoras and Fang [25], Cortell [26] and Khan at al. [27] also investigated these aspects of boundary layer flow. Mushtaq et al. [28] did a numerical study of non-inear radiative heat transfer of nanofluid. Three-dimensional flow of nanofluid over a non-linearly stretching sheet investigated by Khan et al. [29]. Recently, Mahanthesh et al. [30] studied non-linear radiative heat transfer in magnetohydrodynamic 3D flow of water based nanofluid over a non-linearly stretching sheet with convective boundary conditions.

Study of effects of heat source/sink in heat transfer is important in many physical problems. However, the accurate mathematical modeling of internal heat source/sink is seems to be difficult. Heat transfer effects over a nonlinearly stretching sheet with variable wall temperature and non-uniform heat source discussed by Nandeppanavar et al. [31]. Gireesha et al. [32] examined radiative Hall effects on boundary layer flow past a non-isothermal stretching surface with non-uniform heat source/sink and fluid-particle suspension embedded in porous medium. Mabood et al. [33], Pal and Mandal [34] and Raju et al. [35] also studied the non-uniform heat source/sink for boundary layer flow.

The thermophoresis phenomenon in boundary layer flow has great importance. This phenomenon causes due to small particle,that flow away from the sheet. It has applications in gas particle trajectories form combustion devices and turbine blades etc. Rashad [36] studied influence of radiation on magnetohydrodynamic free convection over a vertical flat plate with thermophoretic deposition of particles embedded in porous media. Noor et al. [37] discussed Heat and mass transfer effects of thermophoretic MHD flow in the presence of heat source/sink over an inclined radiate isothermal permeable surface. Thermophoretic MHD slip flow over a permeable surface with variable fluid properties examined by Das et al. [38]. Immaculate et al. [39] investigated the effects of thermophoretic particle deposition on fully developed MHD mixed convective flow in a vertical channel with thermal-diffusion and diffusion-thermo effects. In recent times, transient thermophoretic particle deposition on forced convective heat and mass transfer flow due to a rotating disk was studied by Alam et al. [40].

Finally, study of chemical reaction in boundary layer flow has great significance in chemical engineering industries, such as polymer production and food processing. Its applications are energy transfer in a wet cooling water, desert cooler etc. Bhattacharyya [41] examined dual solutions in boundary layer stagnation-point flow and mass transfer effect with chemical reaction past a stretching/shrinking sheet. Analytical study of MHD free convective, dissipative boundary layer flow in the presence of thermal radiation, chemical reaction and constant suction past a porous vertical surface studied by Raju et al. [42]. Recently, Krishnamurthy et al. [42] investigated the influence of MHD boundary layer flow and melting heat transfer of Williamon nanofluid in porous medium in chemical reaction.

To our best knowledge, the studied on non-linear thermal radiation with non-uniform heat source/sink in MHD convective boundary layer flow in the presence of thermophoretic velocity and chemical reaction has never been investigated till date. Aim of this investigation is to exted the work of Mahanthesh et al. [30]. We have investigated the thermophoretic effect on non-linear radiative heat transfer with convective boundary conditions over a non-linearly stretching sheet. The governing equations are cracked numerically. Results obtained are found in excellent agreement with the literature available. Results obtained are shown through particular graphs and tables for appropriate parameters

## 2. MATHEMATICAL FORMULATION

We have considered incompressible electrically conducting non-linear radiative three-dimensional boundary layer flow along the heated convective stretching sheet with non-uniform heat source (see fig. 1).

The fluid flow along $x$ direction with velocity $u_w = a(x+y)^n$ and along $y$ direction with velocity $v_w = b(x+y)^n$ respectively, where $a$, $b$ are constant and $n > 0$. $T_w$ and $C_w$ are the temperature and concentration at the walls respectively. $T_c$ is the temperature due to convective heating process with heat transfer coefficient $h_c$. $T_\infty$ and $C_\infty$ are the ambient temperature and ambient concentration near the walls correspondingly. In $z$ direction we have applied variable magnetic field as $B = B_0(x+y)^{\frac{n-1}{2}}$, where, $B_0$ is a constant.

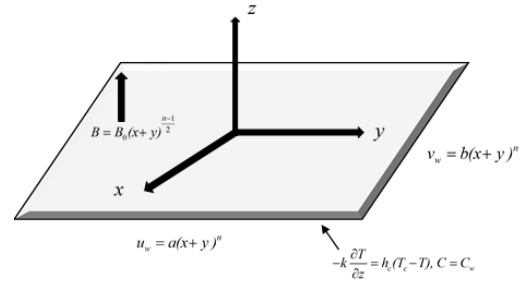

**Fig. 1. Physical modal of the problem.**

Using above boundary layer approximations, equations of momentum, equation of energy and the concentration eqs. are as follows (see refs. [30])

$$\frac{\partial u}{\partial x}+\frac{\partial v}{\partial y}+\frac{\partial w}{\partial z}=0 \quad (1)$$

$$u\frac{\partial u}{\partial x}+v\frac{\partial u}{\partial y}+w\frac{\partial u}{\partial z}=\alpha\frac{\partial^2 u}{\partial z^2}-\frac{\sigma B^2}{\rho}u \quad (2)$$

$$u\frac{\partial v}{\partial x}+v\frac{\partial v}{\partial y}+w\frac{\partial v}{\partial z}=\alpha\frac{\partial^2 v}{\partial z^2}-\frac{\sigma B^2}{\rho}v \quad (3)$$

$$u\frac{\partial T}{\partial x}+v\frac{\partial T}{\partial y}+w\frac{\partial T}{\partial z}=\alpha\frac{\partial^2 T}{\partial z^2}-\frac{1}{\rho C_p}\frac{\partial q_r}{\partial z}$$
$$+\frac{D_B K_T}{C_s C_p}\frac{\partial^2 C}{\partial z^2}+\frac{q'''}{\rho C_p} \quad (4)$$

$$u\frac{\partial C}{\partial x}+v\frac{\partial C}{\partial y}+w\frac{\partial C}{\partial z}=D_B\frac{\partial^2 C}{\partial z^2}+\frac{D_B K_T}{T_m}\frac{\partial^2 T}{\partial z^2}$$
$$-\frac{\partial V_T}{\partial z}(C-C_\infty)-K_r(C-C_\infty) \quad (5)$$

Where, $u$, $v$ and $w$ are the velocity components in $x$, $y$ and $z$ directions respectively. B is the magnetic field, $\mu$ is dynamic viscosity, $\rho$ is density, $\sigma$ is electrical conductivity, $C_p$ is specific heat at constant pressure, T and C are the temperature and concentration respectively. $T_m$ is fluid mean temperature, $C_s$ is concentration susceptibility, $K_T$ is thermal diffusion coefficient, $K_r$ is the chemical reaction coefficient and $D_B$ is species diffusion coefficient.

The space and temperature dependent non-uniform heat source/sink [35] is as follows:

$$q'''=\frac{k u_w}{(x+y)\nu}\left[A^*(T_c-T_\infty)f'(\eta)+B^*(T-T_\infty)\right] \quad (6)$$

Where, k is the thermal conductivity, $\nu$ is the kinematic fluid viscosity, $A^*$ and $B^*$ are the space and temperature dependent heat source/sink parameter.

Thermophoretic velocity $V_T$ is defined is as follows

$$V_T=-\frac{k_V \nu}{T_r}\frac{\partial T}{\partial z} \quad (7)$$

Here, $T_r$ is the reference temperature, $k_V$ is the thermophoretic coefficient [37,38] is given by

$$k_V=\frac{2c_b(\lambda_g/\lambda_p+c_t k_n)(c_1+c_2 e^{-c_3/k_n})}{(1+3c_m k_n)(1+2\lambda_g/\lambda_p+2c_t k_n)} \quad (8)$$

Also where $c_1, c_2, c_3, c_m, c_b, c_t$ are constants and $\lambda_g, \lambda_p$ are thermal conductivity of the fluid and diffused particles correspondingly and $k_n$ is the Knudsen number.

The radiative heat flux $q_r$ by Rosseland approximations is

$$q_r=-\frac{4\sigma^*}{3k_m}\frac{\partial T^4}{\partial z}. \quad (9)$$

Here, $\sigma^*$ is the Stefan-Boltzmann constant and $k_m$ is the mean absorption coefficient. In this study we have considered that radiation is optically thick. Also it is clear that eq. (9) is non-linear in terms of T. Hence by Taylor's theorem $T^4$ can be expressed is

$$T^4=T_\infty^4+4T_\infty^3(T-T_\infty)+6T_\infty^2(T-T_\infty)^2+.... \quad (10)$$

Neglecting higher order terms, then we have

$$T^4=4T_\infty^3-3T_\infty^4. \quad (11)$$

Using eq. (9) and (10) becomes the converted form

$$\frac{\partial q_r}{\partial z}=-\frac{16\sigma^* T_\infty^3}{3k_m}\frac{\partial^2 T}{\partial z^2}. \quad (12)$$

By using eq. (12), eq. (4) becomes

$$u\frac{\partial T}{\partial x}+v\frac{\partial T}{\partial y}+w\frac{\partial T}{\partial z}=\alpha\frac{\partial^2 T}{\partial z^2}+\frac{16\sigma^* T_\infty^3}{3\rho C_p k_m}\frac{\partial^2 T}{\partial z^2}$$
$$+\frac{D_B K_T}{C_s C_p}\frac{\partial^2 C}{\partial z^2}+\frac{q'''}{\rho C_p}. \quad (13)$$

Equation (13) is called linear radiative heat transfer equation.

To find the non-linear thermal radiation, equation (9) can be re-written as:

$$q_r=-\frac{4\sigma^*}{3k_m}\frac{\partial T^4}{\partial z}=-\frac{16\sigma^*}{3k_m}T^3\frac{\partial T}{\partial z}$$

$$\frac{\partial q_r}{\partial z}=-\frac{16\sigma^*}{3k_m}\frac{\partial}{\partial z}\left(T^3\frac{\partial T}{\partial z}\right). \quad (14)$$

Using equation (14), equation (4) becomes

$$u\frac{\partial T}{\partial x}+v\frac{\partial T}{\partial y}+w\frac{\partial T}{\partial z}=\alpha\frac{\partial^2 T}{\partial z^2}$$
$$+\frac{16\sigma^*}{3\rho C_p k_m}\frac{\partial}{\partial z}\left(T^3\frac{\partial T}{\partial z}\right)+\frac{D_B K_T}{C_s C_p}\frac{\partial^2 C}{\partial z^2}+\frac{q'''}{\rho C_p} \quad (15)$$

Equation (15) is called non-linear thermal radiation heat transfer equation.

Subject to boundary conditions:

$$u=u_w, \quad v=v_w, \quad w=0,$$
$$-k\frac{\partial T}{\partial z}=h_c(T_c-T), \quad C=C_w, \text{ at } z=0$$
$$u\to 0, \quad v\to 0, \quad T\to T_\infty, \quad C\to C_\infty \text{ as } z\to\infty \quad (16)$$

Now introducing the suitable similarity transformation (see refs. [30])

$$u=a(x+y)^n f'(\eta), \quad v=a(x+y)^n g'(\eta)$$

$$w=-\sqrt{a\nu}(x+y)^{\frac{n-1}{2}}\left[\begin{array}{l}\frac{n+1}{2}(f(\eta)+g(\eta))\\+\frac{n-1}{2}\eta(f'(\eta)+g'(\eta))\end{array}\right],$$

$$T-T_\infty=(T_c-T_\infty)\theta(\eta),$$

$$C-C_\infty=(C_w-C_\infty)\psi(\eta), \quad \eta=\sqrt{\frac{a}{\nu}}(x+y)^{\frac{n-1}{2}}z. \quad (17)$$

With the help of above similarity transformation equations (2), (3), (13), (15) and (5) reduces into eqs. (18-22) and eq. (16) reduces into eq. (23).

$$f'''+\left(\frac{n+1}{2}\right)(f+g)f''-n(f'+g')f'-Mf'=0 \quad (18)$$

$$g'''+\left(\frac{n+1}{2}\right)(f+g)g''-n(f'+g')g'-Mg'=0 \quad (19)$$

$$\frac{1}{Pr}(1+R)\theta'' + \left(\frac{n+1}{2}\right)(f+g)\theta' + Df\,\psi''$$
$$+ \frac{1}{Pr}(A^*f' + B^*\theta) = 0 \tag{20}$$

$$\frac{1}{Pr}\left[1 + R\{(\theta_p - 1)\theta + 1\}^3\right]\theta'' + \frac{3R}{Pr}(\theta_p - 1)$$
$$\left[(\theta_p - 1)\theta + 1\right]^2 \theta'^2 + \left(\frac{n+1}{2}\right)(f+g)\theta' \tag{21}$$
$$+ Df\,\psi'' + \frac{1}{Pr}(A^*f' + B^*\theta) = 0$$

$$\psi'' + Sc\left(\frac{n+1}{2}\right)(f+g)\psi' + Sc\,Sr$$
$$-\tau Sc(\psi\theta'' + \psi'\theta') - Sc\,K_R\psi = 0 \tag{22}$$

subject boundary conditions
$f' = 1$, $g' = c$, $f = 0$, $g = 0$, $\theta' = Bi(\theta - 1)$, $\psi = 1$ at $\eta = 0$.
$f' \to 0$, $g' \to 0$, $\theta \to 0$, $\psi \to 0$, as $\eta \to \infty$. (23)

Where M is the magnetic field parameter, R is the radiation parameter, Pr is the Prandtl number, $Df$ is the Dufour number, $A^*$ and $B^*$ is the non-uniform heat source/sink parameters, $\theta_p$ is the temperature ratio parameter, Sc is the Schmidt number, Sr is the Soret number, $\tau$ is the thermophoretic parameter, $K_R$ is the chemical reaction parameter, $c$ is the stretching ratio parameter and $Bi$ is the Biot number. We take $h_c = c^*/\sqrt{x^{1-n}}$ for similarity solution, where $c^*$ is the constant. The parameters are defined as:

$$M = \frac{\sigma B_0^2}{\rho a}, \quad R = \frac{16\sigma^* T_\infty^3}{3k\,k_m}, \quad Pr = \frac{\rho C_p \nu}{k},$$
$$Df = \frac{D_B K_T(C_w - C_\infty)}{\nu C_s C_p(T_c - T_\infty)}, \quad \theta_p = \frac{T_c}{T_\infty}, \quad Sc = \frac{\nu}{D_B},$$
$$Sr = \frac{D_B K_T(T_c - T_\infty)}{T_m \nu(C_w - C_\infty)}, \quad \tau = -\frac{k_V(T_c - T_w)}{T_r},$$
$$K_R = \frac{K_r(x+y)}{u_w}, \quad c = \frac{b}{a}, \quad Bi = \frac{c^*}{k}\sqrt{\frac{\nu}{a}}. \tag{24}$$

Skin friction coefficient, local Nusselt number and Sherwood number is defined as follows (see refs. [30,35])

$$C_{fx} = \frac{\tau_{zx}}{\rho u_w^2}, \quad C_{fy} = \frac{\tau_{zy}}{\rho v_w^2},$$
$$Nu = \frac{(x+y)q_w}{k(T_c - T_\infty)}, \quad Sh = \frac{(x+y)m_w}{D_B(C_w - C_\infty)}, \tag{25}$$

Here, $\tau_{zx}$ and $\tau_{zy}$ are shear stresses along the $x$ and $y$ directions at the wall, $q_w$ and $m_w$ are the heat flux and mass flux respectively at the wall.

$$\tau_{zx} = \mu\left(\frac{\partial u}{\partial z} + \frac{\partial w}{\partial x}\right)_{z=0}, \quad \tau_{zy} = \mu\left(\frac{\partial v}{\partial z} + \frac{\partial w}{\partial y}\right)_{z=0},$$
$$q_w = -k\left(\frac{\partial T}{\partial z}\right)_{z=0}, \quad m_w = -D_B\left(\frac{\partial C}{\partial z}\right)_{z=0}. \tag{26}$$

By using similarity transformation equation (25) becomes

$$\sqrt{Re_x}\,C_{fx} = f''(0), \quad \sqrt{Re_y}\,C_{fy} = g''(0),$$
$$\frac{Nu}{\sqrt{Re_x}} = -\theta'(0), \quad \frac{Sh}{\sqrt{Re_x}} = -\psi'(0). \tag{27}$$

Where, $Re_x = \frac{u_w(x+y)}{\nu}$ and $Re_y = \frac{v_w(x+y)}{\nu}$ are the local Reynolds number along the $x$ and $y$ directions correspondingly.

## 3. METHOD OF SOLUTION

To solve the equations (18)-(22) under the boundary conditions (23), we have converted these equation from BVP to IVP. These coupled non-linear equations are solved by fourth-fifth order Runge-Kutta-Fehlberg scheme with shooting techniques in MATLAB. We use hit and trial method for guess the initial values of $f''(0)$, $g''(0)$, $\theta'(0)$ and $\psi'(0)$. Secant method is used to find better approximations for these values. The RKF45 or Runge-Kutta-Fehlberg method is a well-tested method to find out the solutions of initial value problems. In this method it is very necessary to choose the exact finite value of $\eta_\infty$. For this particular study, we have chosen $\eta_\infty$ is $\eta_8$. The step size is taken $\Delta\eta = 0.001$ for numerical calculations, the process is repeated till that we get the desired convergence accuracy of $10^{-6}$.

The formula of RKF-45 method is written as

$$y_{m+1} = y_m + h\left(\frac{25}{216}k_0 + \frac{1408}{2565}k_2 + \frac{2197}{4104}k_3 - \frac{1}{5}k_4\right), \tag{28}$$

$$y_{m+1} = y_m + h\left(\begin{array}{c}\frac{16}{135}k_0 + \frac{6656}{12825}k_2 + \frac{28561}{56430}k_3 \\ -\frac{9}{50}k_4 + \frac{2}{25}k_5\end{array}\right), \tag{29}$$

Equations (28) and (29) are fourth and fifth order Runge-Kutta scheme respectively.

$$k_0 = f(x_m, y_m),$$
$$k_1 = f\left(x_m + \frac{h}{4}, y_m + \frac{hk_0}{4}\right),$$
$$k_2 = f\left(x_m + \frac{3}{8}h, y_m + \left(\frac{3}{32}k_0 + \frac{9}{32}k_1\right)h\right),$$
$$k_3 = f\left(x_m + \frac{12}{13}h, y_m + \left(\begin{array}{c}\frac{1932}{2197}k_0 - \frac{7200}{2197}k_1 \\ + \frac{7296}{2197}k_2\end{array}\right)h\right),$$
$$k_4 = f\left(x_m + h, y_m + \left(\begin{array}{c}\frac{439}{216}k_0 - 8k_1 + \frac{3680}{513}k_2 \\ -\frac{845}{4104}k_3\end{array}\right)h\right),$$

$$k_5 = f\left(x_m + \frac{h}{2}, y_m + \left(\begin{array}{c}-\frac{8}{27}k_0 + 2k_1 - \frac{3544}{2565}k_2 \\ +\frac{1859}{4104}k_3 - \frac{11}{40}k_4\end{array}\right)h\right).$$

## 4. RESULT AND DISCUSSION

A study incompressible electrically conducting non-linear radiative boundary layer three-dimensional flow along the heated convective stretching sheet with non-uniform heat source is studied numerically. Mainly the effects of non-linear thermal radiation, Soret and Dufour effects, chemical reaction effects and thermophoretic effects are investigated. The present study is analyze the impact of various parameters on velocity and temperature.

Table 1 shows the comparison between Khan et al. [29] and Mahanthesh et al. [30] with the results obtained in present investigation. It is clearly seen form the table that the present results are very well in agreement with the results obtained by Khan et al. [29] and Mahanthesh et al. [30]. This is also shows the accuracy of the present result.

Figure 2 and 3 shows the variation of stretching ratio parameter c on both axial and transverse velocities. It is very clear form these figure that an increase in stretching ration parameter c, displays the reduction of boundary layer thickness in axial velocity, whereas exactly opposite results shown for transverse velocity. Because for large values of $c = b/a$, transverse velocity shows enhancement due to constant $b$, because $b$ is stretching parameter along $y$-direction and axial velocity decline in $x$-direction, due to constant stretching parameter a. The similar results were obtained by Mahantesh et al. [30].

Figure 4 and 5 depicts the variation of magnetic field parameter on both axial and transverse velocities. Magnetic field parameter creates a resistance force, called Lorentz force. The Lorentz force is well known to work in opposite direction to the fluid flow. Hence enhancement in magnetic field leads declined in velocities profiles. Figure 6 displays that for large values of power law index $n$, velocities profiles decreases, hence boundary layer thickness reduces. Figures 7 and 8 describes the influence of stretching ratio parameter c on temperature and concentration profiles. When the stretching ration parameter c increase, leads to intensity of the cooler fluid to hotter fluid from ambient fluid close to the surface is increases. Hence the thermal boundary layer thickness decreases as well as concentration profile shows reduction too. Figure 9 and 10 indicates the effects of magnetic field parameter M on temperature profile and concentration profile. Due to Lorentz force it is clearly seen that for higher values of magnetic parameter M, both temperature profile and concentration increases.

It is observed form figure 11 that as power law index $n$ increases, temperature profile and concentration profile decreases. Figure 12 shows the effects of Biot number Bi on temperature profile. This figure indicates that higher Biot number increases flow of temperature profile, this is due to increment in the temperature difference for every increasing value of Biot number. In figure 13 we have studied the effect of Dufour number on temperature and concentration distributions. Increasing Dufour number causes the heat of fluid flow increases. Hence for every increasing value of Dufour number both temperature and concentration profiles displays enhancement. Figure 14 and 15 depicts the influence of Soret number on temperature profile and concentration profile. Soret number illustrate the impact of temperature gradients, this number displays the dual effects in both profiles. An increment in Soret number, the thermal boundary layer thickness decreases, while the solute boundary layer thickness shows growth.

Figures 16 and 17 illustrates the reduction of thermal boundary layer thickness and solute boundary layer thickness for every large value of Prandtl number Pr. It is observed form the figures 18 and 19 that for large value of Schmidt number Sc, temperature profile increases, whereas concentration profile shows reduction. The influence of non-uniform heat source/sink parameters A* and B* are indicates in figures 20-23. It is seen that the increases in the value of A* and B*, shows the enhancement of thermal boundary layer thickness and concentration boundary layer thickness. This is due to that the presence of heat source release the heat or energy to the flow, this energy helps to improve the thermal and solute boundary layers thickness. Similar results were obtained by Raju et al. [35]

Figure 24 shows the effects of radiation parameter R in temperature profile. This figure also depicts comparison between linear thermal radiation and non-linear thermal radiation on thermal boundary layer. It can be seen clearly that when we increase linear and non-linear thermal radiation, the thermal boundary layer thickness increases. This is because the radiation gives energy to the flow and helps to improve the heats of the fluid. Hence the thermal boundary layer thickness increases as we increase radiation parameter R. Also noted that the increment of non-linear thermal radiation in thermal boundary layer is much higher than linear thermal radiation. Similar results are obtained by Mahanthesh et al. [30] and Mushtaq et al. [28]. Therefore we can say that non-linear thermal radiation is more effective than linear thermal radiation, these results will helps to improve the capabilities of nuclear power plants and gas turbines.

Figures 25 and 26 depicts the effects of temperature ratio parameter $\theta_p = T_c / T_\infty$ on temperature profile and concentration profile. As the temperature ratio parameter increases, both temperature profile and concentration profile increases. Because when we increases the stretching sheet temperature $T_c$, normally the temperature ratio parameter shows increment. Hence thermal boundary layer thickness and concentration boundary layer thickness increases with temperature ratio parameter $\theta_p$. These outcomes are in very good agreement with Mahanthesh et al. [30] for viscous fluids.

Figure 27 illustrate the variation of thermophoretic parameter $\tau$ on concentration profile. From the figure it is observed that for increasing values of thermophoretic parameter $\tau$, concentration boundary layer thickness reduces. These results are in very well agreement with Das et al. [38]. Figure 28 shows that chemical reaction parameter $K_R$ effects on the concentration distribution. It is observed that enhancement in chemical reaction parameter $K_R$, leads to reduce the concentration boundary layer thickness. It is due to consumption of chemicals take place during chemical reaction..

Table 2 shows the numerical results of skin friction coefficient along *x* and *y* directions. In this table we obtained the values of $f''(0)$ and $g''(0)$ for the variation of power law index *n*, stretching ratio parameter *c* and magnetic field parameter M. The negative values indicates that the sheet applies a drag force on the fluid flow. It is observed from the table that skin friction coefficient along both the directions shows reduction when we increase power law index *n*, magnetic field parameter M and stretching ratio parameter *c*.

Table 3 and 4 describes the numerical values of Nusselt number and Sherwood number from *n* = 1 and *n* = 3, respectively for different values of c, M, Pr, R, Bi, $\theta_p$, A*, B*, Df, Sr, Sc, $\tau$ and $K_R$. It is clear seen that for increasing values of c, Pr, Bi and Sr, Nusselt number increases for both the values of *n*, while Nusselt number decreases for increasing values of M, R, $\theta_p$, A*, B*, Df, Sc, $\tau$ and $K_R$. On the other hand Sherwood number shows enhancement for increasing values of c, Pr, Bi, Sc, $\tau$ and $K_R$, whereas Sherwood number decreases with the increasing values of M, R, $\theta_p$, A*, B*, Df and Sr.

## 5. CONCLUSION

The present paper aimed to study the effects of MHD boundary layer flow of non-linear thermal radiation with convective boundary conditions and non-uniform heat source/sink in the presence of thermophoretic velocity and chemical reaction. The results obtained are as follows.

--Axial velocity profile decreases with stretching ratio parameter, while transverse velocity profile increases as we increase stretching ratio parameter *c*.

--An increment in magnetic field parameter M and power law index *n*, leads to reduce both velocity profiles.

--An increase in magnetic field parameter M causes increase temperature and concentration profile increases, whereas for higher value of power law index *n*, both profile shows reduction.

--Positive values of heat source/sink coefficients A* and B*, enhanced heat transfer rate.

Non-linear thermal radiation shows more effectiveness than linear thermal radiation. Hence non-linear thermal radiation helps us to improve over nuclear power plants.

--An increment of therophoretic parameter $\tau$ and chemical reaction parameter $K_R$, the thickness of concentration boundary layer decreases.

**Conflict of interest:** There is no conflict of interest for publishing this paper.

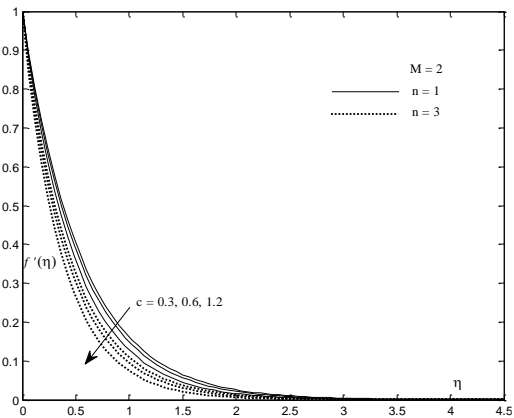

**Fig.2. Effect of stretching ratio parameter c on axial velocity.**

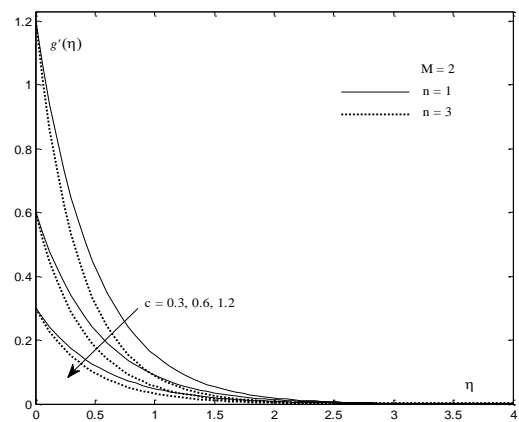

**Fig.3. Effect of stretching ratio parameter c on transverse velocity.**

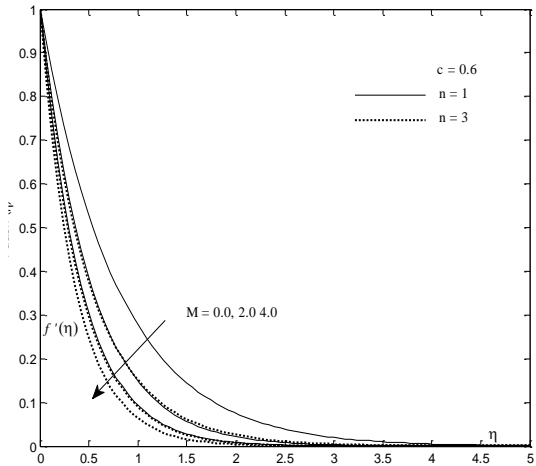
Fig.4. Effect of magnetic field parameter M on axial velocity.

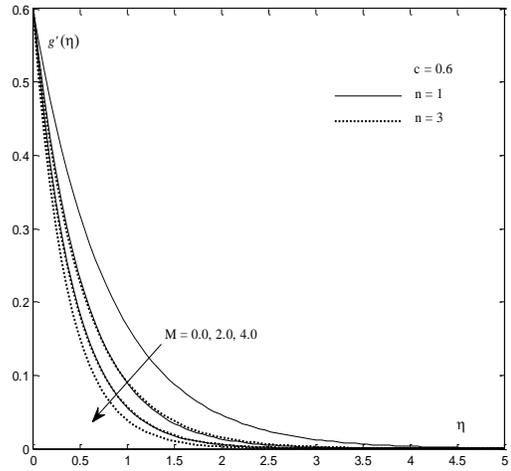
Fig.5. Effect of magnetic field parameter M on transverse velocity.

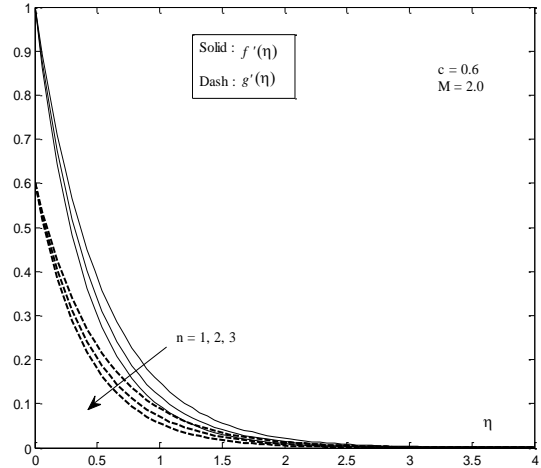
Fig.6. Effect of power law index *n* on axial velocity and transverse velocity.

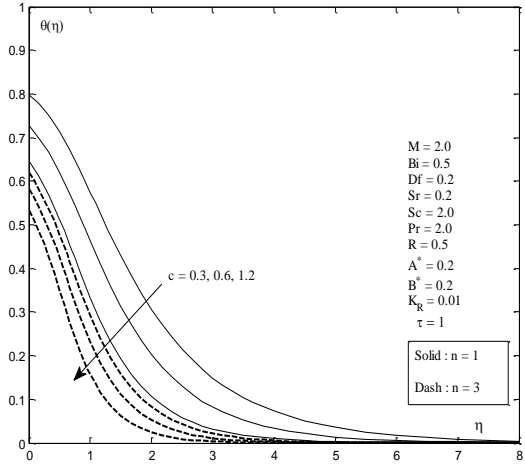
Fig.7. Effect of stretching ratio parameter c on temperature.

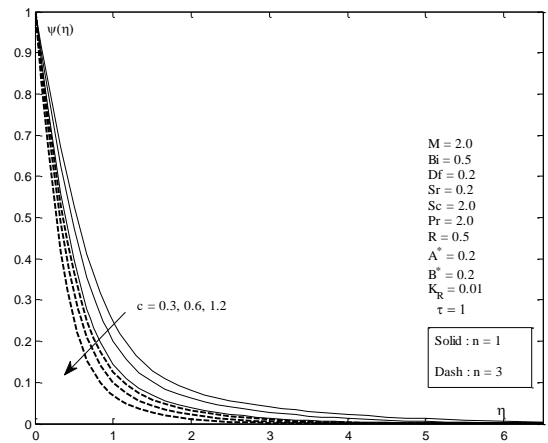
Fig.8. Effect of stretching ratio parameter c on concentration.

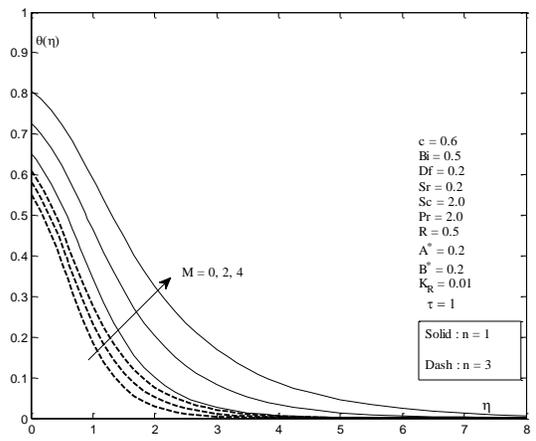
Fig.9. Effect of magnetic field parameter M on temperature.

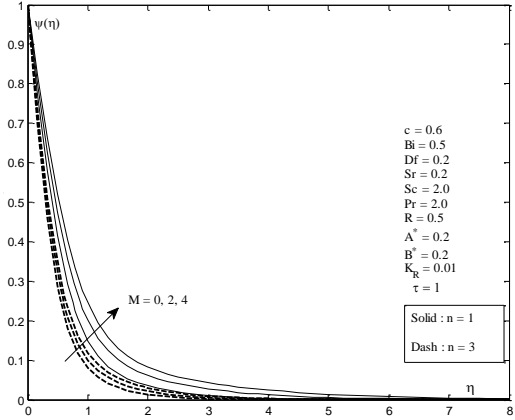

**Fig.10. Effect of magnetic field parameter M on concentration.**

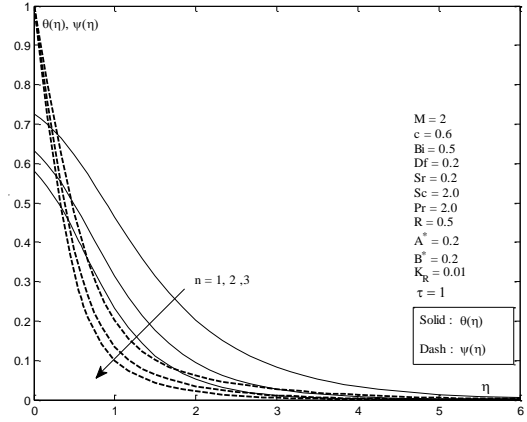

**Fig.11. Effect of power law index *n* on temperature and concentration.**

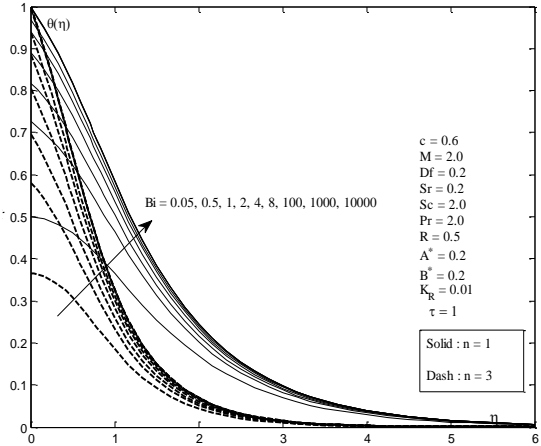

**Fig.12. Effect of Biot number *Bi* on temperature.**

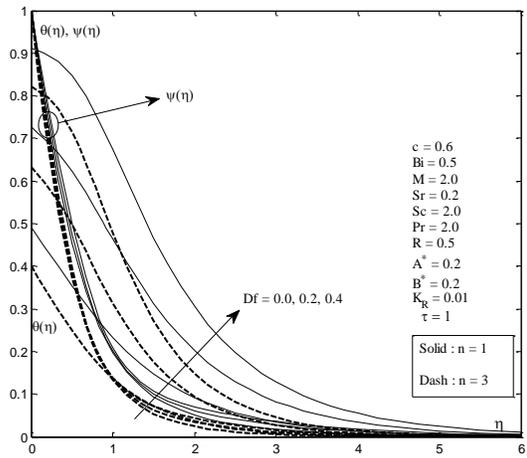

**Fig.13. Effect of Dufour number *Df* on temperature and concentration.**

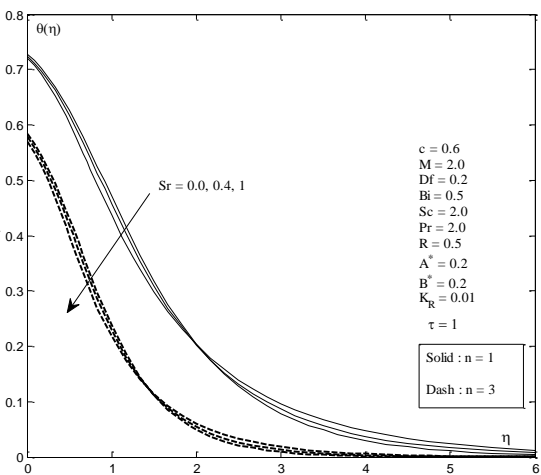

**Fig.14. Effect of Soret number *Sr* on temperature.**

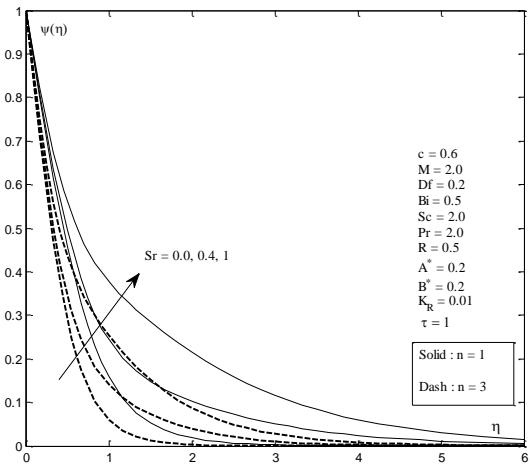

**Fig.15. Effect of Soret number *Sr* on concentration.**

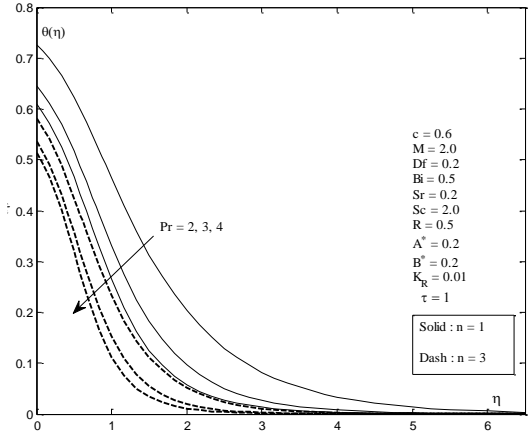
**Fig.16. Effect of Prandtl number *Pr* on temperature.**

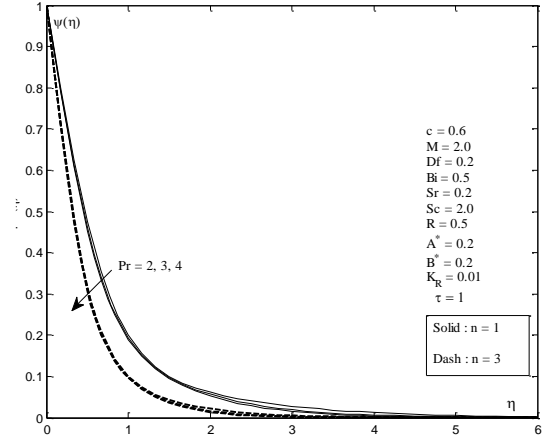
**Fig.17. Effect of Prandtl number *Pr* on concentration.**

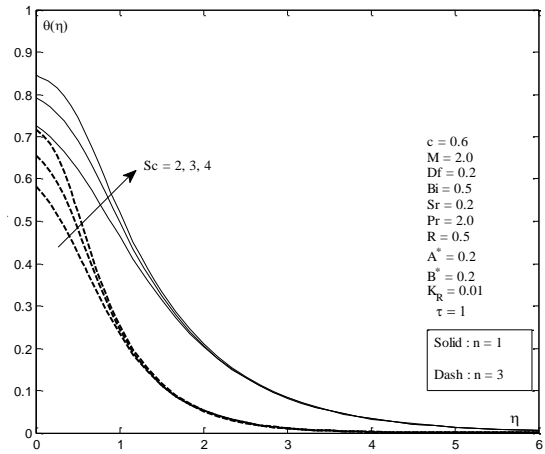
**Fig.18. Effect of Schmidt number *Sc* on temperature.**

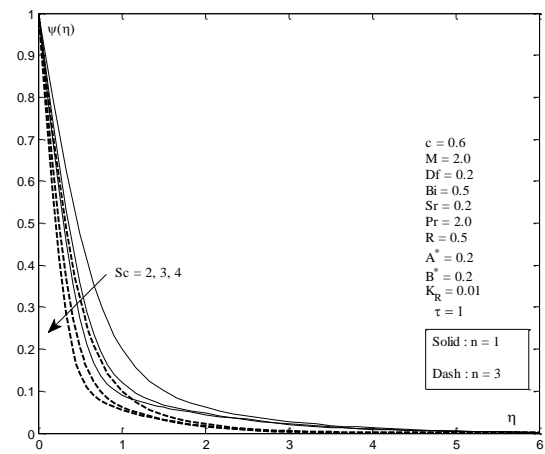
**Fig.19. Effect of Schmidt number *Sc* on concentration.**

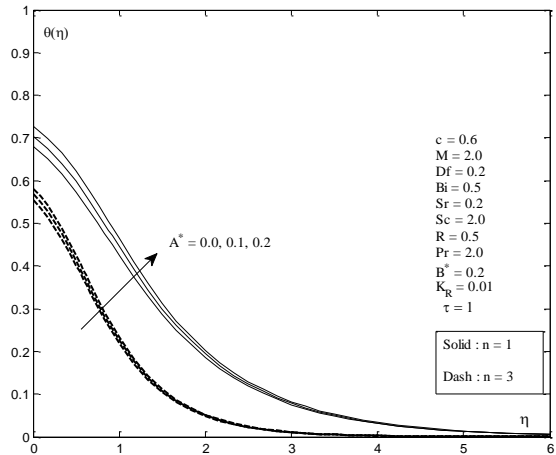
**Fig.20. Effect of heat source/sink parameter A* on temperature.**

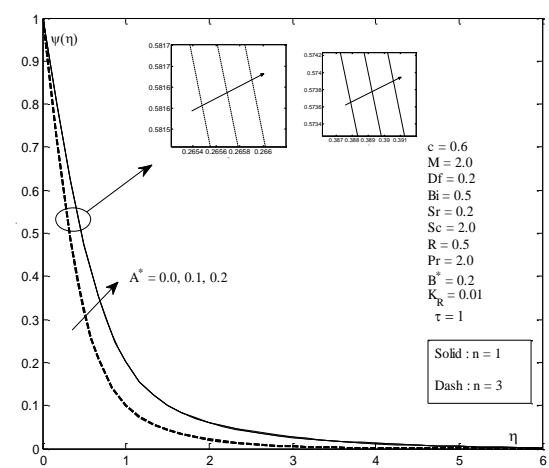
**Fig.21. Effect of heat source/sink parameter A* on concentration.**

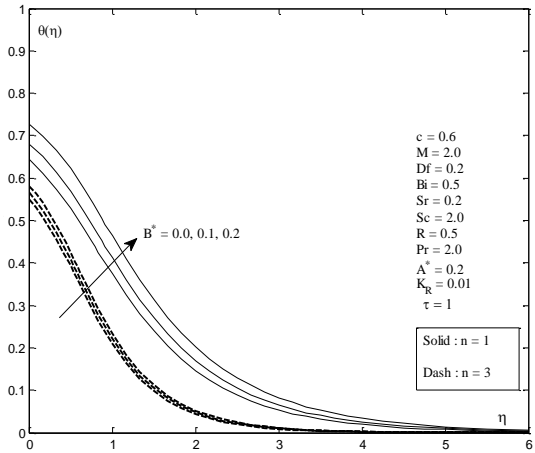

**Fig.22. Effect of heat source/sink parameter $B^*$ on temperature.**

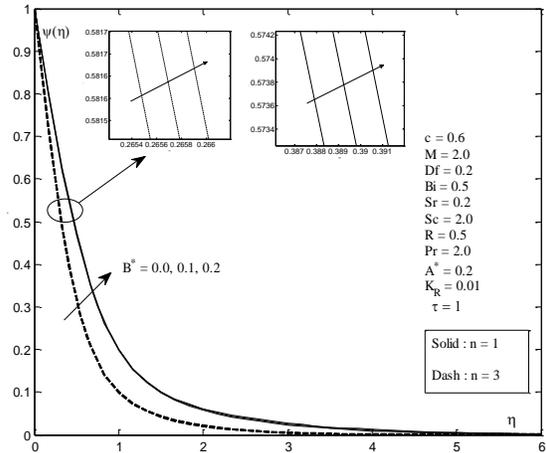

**Fig.23. Effect of heat source/sink parameter $B^*$ on concentration.**

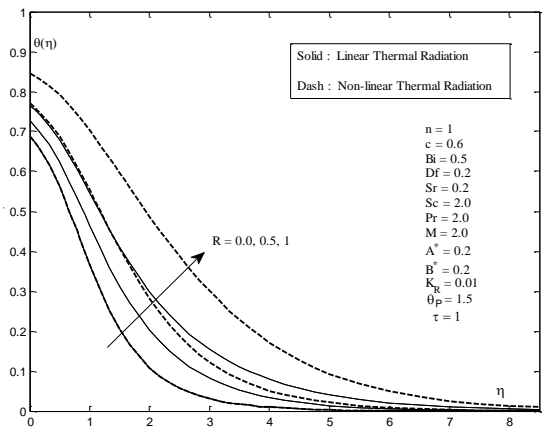

**Fig.24 Effect of radiation parameter R on temperature.**

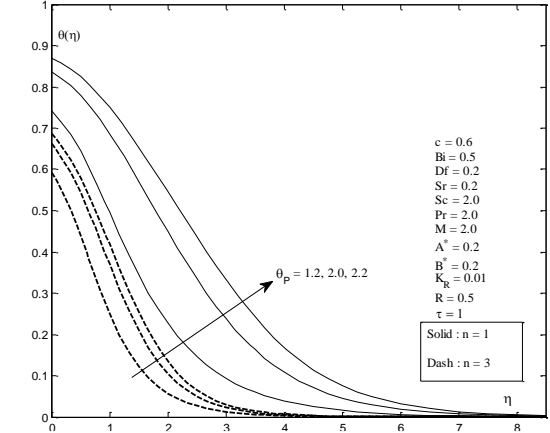

**Fig.25 Effect of temperature ratio parameter $\theta_p$ on temperature.**

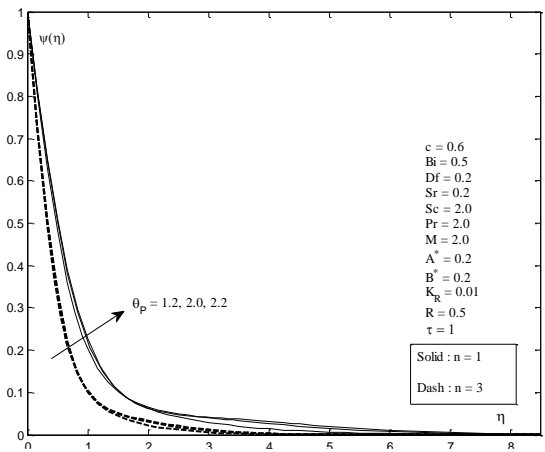

**Fig.26 Effect of temperature ratio parameter $\theta_p$ on concentration.**

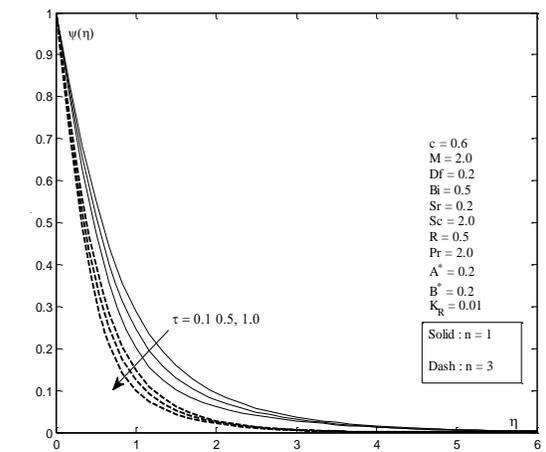

**Fig.27 Effect of thermophoretic parameter $\tau$ on concentration.**

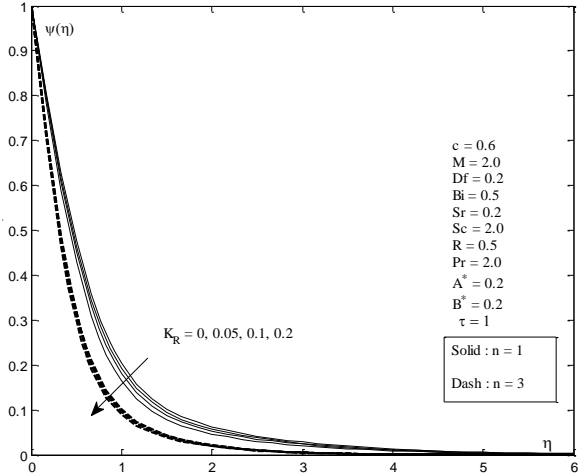

**Fig.28. Effect of chemical reaction parameter $K_R$ on concentration**

**Table 1. Comparison of the values of $f''(0)$ and $g''(0)$ when M = 0.**

| n | c | Junaid et al. | | Mahanthesh et al. | | Present Study | |
|---|---|---|---|---|---|---|---|
| | | $f''(0)$ | $g''(0)$ | $f''(0)$ | $g''(0)$ | $f''(0)$ | $g''(0)$ |
| 1 | 0 | -1 | 0 | -1 | 0 | -1 | 0 |
| 1 | 0.5 | -1.224745 | -0.612372 | -1.22474 | -0.61237 | -1.229491946 | -0.614934948 |
| 1 | 1 | -1.414214 | -1.414214 | -1.41421 | -1.41421 | -1.419135111 | -1.419135231 |
| 3 | 0 | -1.624356 | 0 | -1.62436 | 0 | -1.628751125 | 0 |
| 3 | 0.5 | -1.989422 | -0.994711 | -1.98952 | -0.99471 | -1.993996413 | -0.997569903 |
| 3 | 1 | -2.297186 | -2.297186 | -2.29719 | -2.29712 | -2.301331346 | -2.301341846 |

**Table 2. Numerical values of Skin friction coefficient along *x* and *y* directions.**

| n | c | M | $\text{Re}_x^{1/2} C_{fx}$ | $\text{Re}_x^{1/2} C_{fy}$ |
|---|---|---|---|---|
| 1 | 0.5 | 2 | -1.870828693 | -0.935414347 |
| | 1 | 2 | -2.000000000 | -2.000000000 |
| | 1.5 | 2 | -2.121320344 | -3.181980515 |
| | 2 | 2 | -2.236067977 | -4.472135955 |
| | 0.5 | 0 | -1.224744874 | -0.612372437 |
| | 0.5 | 2 | -1.870828693 | -0.935414347 |
| | 0.5 | 4 | -2.345207880 | -1.172603940 |
| | 0.5 | 6 | -2.738612788 | -1.369306394 |
| 3 | 0.5 | 2 | -2.444438554 | -1.222219277 |
| | 1 | 2 | -2.701221616 | -2.701221616 |
| | 1.5 | 2 | -2.935539579 | -4.403309369 |
| | 2 | 2 | -3.152432993 | -6.304865986 |
| | 0.5 | 0 | -1.989427446 | -0.994713723 |
| | 0.5 | 2 | -2.444438554 | -1.222219277 |
| | 0.5 | 4 | -2.825319062 | -1.412659531 |
| | 0.5 | 6 | -3.160121685 | -1.580060842 |

**Table 3. Numerical values of Nusselt number and Sherwood number when *n* = 1.**

| c | M | Pr | R | Bi | $\theta_p$ | A* | B* | Df | Sr | Sc | $\tau$ | $K_R$ | Nu | Sh |
|---|---|----|---|----|------------|----|----|----|----|----|--------|-------|----|----|
| 0.5 | 2 | 2 | 0.5 | 0.5 | 1.2 | 0.2 | 0.2 | 0.2 | 0.2 | 2.0 | 1 | 0.01 | 0.119225036 | 1.127128726 |
| 0.7 |   |   |   |   |   |   |   |   |   |   |   |   | 0.138489887 | 1.227046461 |
| 1.5 |   |   |   |   |   |   |   |   |   |   |   |   | 0.184709814 | 1.548994748 |
| 0.5 | 0 |   |   |   |   |   |   |   |   |   |   |   | 0.162580083 | 1.309481228 |
|   | 2 |   |   |   |   |   |   |   |   |   |   |   | 0.119225036 | 1.127128726 |
|   | 4 |   |   |   |   |   |   |   |   |   |   |   | 0.068864379 | 0.977958978 |
|   | 2 | 2 |   |   |   |   |   |   |   |   |   |   | 0.119225036 | 1.127128726 |
|   |   | 3 |   |   |   |   |   |   |   |   |   |   | 0.165257721 | 1.183818678 |
|   |   | 4 |   |   |   |   |   |   |   |   |   |   | 0.185658659 | 1.208128425 |
|   |   | 2 | 0.0 |   |   |   |   |   |   |   |   |   | 0.147373656 | 1.156000953 |
|   |   |   | 0.5 |   |   |   |   |   |   |   |   |   | 0.119225036 | 1.127128726 |
|   |   |   | 1 |   |   |   |   |   |   |   |   |   | 0.091102569 | 1.095400303 |
|   |   |   | 0.5 | 0.5 |   |   |   |   |   |   |   |   | 0.119225036 | 1.127128726 |
|   |   |   |   | 1 |   |   |   |   |   |   |   |   | 0.152801674 | 1.169829752 |
|   |   |   |   | 2 |   |   |   |   |   |   |   |   | 0.177269777 | 1.201145543 |
|   |   |   |   | 0.5 | 1.2 |   |   |   |   |   |   |   | 0.119225036 | 1.127128726 |
|   |   |   |   |   | 2.0 |   |   |   |   |   |   |   | 0.068947980 | 1.070972831 |
|   |   |   |   |   | 2.2 |   |   |   |   |   |   |   | 0.051743956 | 1.051155480 |
|   |   |   |   |   | 1.2 | 0.2 |   |   |   |   |   |   | 0.119225036 | 1.127128726 |
|   |   |   |   |   |   | 0.3 |   |   |   |   |   |   | 0.107397720 | 1.110048919 |
|   |   |   |   |   |   | 0.4 |   |   |   |   |   |   | 0.095613884 | 1.093044715 |
|   |   |   |   |   |   | 0.2 | 0.2 |   |   |   |   |   | 0.119225036 | 1.127128726 |
|   |   |   |   |   |   |   | 0.3 |   |   |   |   |   | 0.078434448 | 1.073272862 |
|   |   |   |   |   |   |   | 0.4 |   |   |   |   |   | 0.015940523 | 0.951652140 |
|   |   |   |   |   |   |   | 0.2 | 0.0 |   |   |   |   | 0.234590750 | 1.283983260 |
|   |   |   |   |   |   |   |   | 0.2 |   |   |   |   | 0.119225036 | 1.127128726 |
|   |   |   |   |   |   |   |   | 0.4 |   |   |   |   | 0.030601476 | 1.012736882 |
|   |   |   |   |   |   |   |   | 0.2 | 0.0 |   |   |   | 0.118595363 | 1.144634768 |
|   |   |   |   |   |   |   |   |   | 0.4 |   |   |   | 0.119809254 | 1.109789134 |
|   |   |   |   |   |   |   |   |   | 1 |   |   |   | 0.121111454 | 1.060104082 |
|   |   |   |   |   |   |   |   |   | 0.2 | 2 |   |   | 0.119225036 | 1.127128726 |
|   |   |   |   |   |   |   |   |   |   | 3 |   |   | 0.088485815 | 1.444256339 |
|   |   |   |   |   |   |   |   |   |   | 4 |   |   | 0.063626533 | 1.696619170 |
|   |   |   |   |   |   |   |   |   |   | 2 | 0.1 |   | 0.131580647 | 0.988519114 |
|   |   |   |   |   |   |   |   |   |   |   | 0.5 |   | 0.125891281 | 1.052831387 |
|   |   |   |   |   |   |   |   |   |   |   | 1 |   | 0.119225036 | 1.127128726 |
|   |   |   |   |   |   |   |   |   |   |   | 1 | 0.05 | 0.115378728 | 1.166132008 |
|   |   |   |   |   |   |   |   |   |   |   |   | 0.1 | 0.110835846 | 1.212039046 |
|   |   |   |   |   |   |   |   |   |   |   |   | 0.2 | 0.102449564 | 1.296423972 |

| c | M | Pr | R | Bi | $\theta_p$ | A* | B* | Df | Sr | Sc | $\tau$ | $K_R$ | Nu | Sh |
|---|---|---|---|---|---|---|---|---|---|---|---|---|---|---|
| 0.5 | 2 | 2 | 0.5 | 0.5 | 1.2 | 0.2 | 0.2 | 0.2 | 0.2 | 2.0 | 1 | 0.01 | 0.198146216 | 1.682553014 |
| 0.7 |   |   |   |   |   |   |   |   |   |   |   |   | 0.209137670 | 1.799166641 |
| 1.5 |   |   |   |   |   |   |   |   |   |   |   |   | 0.238433159 | 2.194211798 |
| 0.5 | 0 |   |   |   |   |   |   |   |   |   |   |   | 0.215342932 | 1.798054864 |
|   | 2 |   |   |   |   |   |   |   |   |   |   |   | 0.198146216 | 1.682553014 |
|   | 4 |   |   |   |   |   |   |   |   |   |   |   | 0.182405128 | 1.588420901 |
|   | 2 | 2 |   |   |   |   |   |   |   |   |   |   | 0.198146216 | 1.682553014 |
|   |   | 3 |   |   |   |   |   |   |   |   |   |   | 0.223603604 | 1.710939942 |
|   |   | 4 |   |   |   |   |   |   |   |   |   |   | 0.235938282 | 1.723320261 |
|   |   | 2 | 0.0 |   |   |   |   |   |   |   |   |   | 0.217058184 | 1.699134011 |
|   |   |   | 0.5 |   |   |   |   |   |   |   |   |   | 0.198146216 | 1.682553014 |
|   |   |   | 1 |   |   |   |   |   |   |   |   |   | 0.180037993 | 1.663707625 |
|   |   |   | 0.5 | 0.5 |   |   |   |   |   |   |   |   | 0.198146216 | 1.682553014 |
|   |   |   |   | 1 |   |   |   |   |   |   |   |   | 0.279014121 | 1.784906455 |
|   |   |   |   | 2 |   |   |   |   |   |   |   |   | 0.348636627 | 1.874038912 |
|   |   |   |   | 0.5 | 1.2 |   |   |   |   |   |   |   | 0.198146216 | 1.682553014 |
|   |   |   |   |   | 2.0 |   |   |   |   |   |   |   | 0.162143427 | 1.642853406 |
|   |   |   |   |   | 2.2 |   |   |   |   |   |   |   | 0.149203320 | 1.628203733 |
|   |   |   |   |   | 1.2 | 0.2 |   |   |   |   |   |   | 0.198146216 | 1.682553014 |
|   |   |   |   |   |   | 0.3 |   |   |   |   |   |   | 0.191359045 | 1.672481609 |
|   |   |   |   |   |   | 0.4 |   |   |   |   |   |   | 0.184583951 | 1.662433192 |
|   |   |   |   |   |   | 0.2 | 0.2 |   |   |   |   |   | 0.198146216 | 1.682553014 |
|   |   |   |   |   |   |   | 0.3 |   |   |   |   |   | 0.187741385 | 1.668343464 |
|   |   |   |   |   |   |   | 0.4 |   |   |   |   |   | 0.175929881 | 1.652291343 |
|   |   |   |   |   |   |   | 0.2 | 0.0 |   |   |   |   | 0.313141149 | 1.841967861 |
|   |   |   |   |   |   |   |   | 0.2 |   |   |   |   | 0.198146216 | 1.682553014 |
|   |   |   |   |   |   |   |   | 0.4 |   |   |   |   | 0.105007605 | 1.560518968 |
|   |   |   |   |   |   |   |   | 0.2 | 0.0 |   |   |   | 0.196648734 | 1.714149447 |
|   |   |   |   |   |   |   |   |   | 0.4 |   |   |   | 0.199652337 | 1.650565450 |
|   |   |   |   |   |   |   |   |   | 1 |   |   |   | 0.204144961 | 1.552858187 |
|   |   |   |   |   |   |   |   |   | 0.2 | 2 |   |   | 0.198146216 | 1.682553014 |
|   |   |   |   |   |   |   |   |   |   | 3 |   |   | 0.163060487 | 2.165844327 |
|   |   |   |   |   |   |   |   |   |   | 4 |   |   | 0.133909225 | 2.560636162 |
|   |   |   |   |   |   |   |   |   |   | 2 | 0.1 |   | 0.215041619 | 1.441994364 |
|   |   |   |   |   |   |   |   |   |   |   | 0.5 |   | 0.207222311 | 1.553375172 |
|   |   |   |   |   |   |   |   |   |   |   | 1 |   | 0.198146216 | 1.682553014 |
|   |   |   |   |   |   |   |   |   |   |   | 1 | 0.05 | 0.196199541 | 1.708810213 |
|   |   |   |   |   |   |   |   |   |   |   |   | 0.1 | 0.193824352 | 1.740807750 |
|   |   |   |   |   |   |   |   |   |   |   |   | 0.2 | 0.189249421 | 1.802335132 |

**Table 4. Numerical values of Nusselt number and Sherwood number when *n* = 3.**